\documentclass[lettersize,journal]{IEEEtran}
\usepackage{amsmath,amsfonts}
\usepackage{algorithm}
\usepackage{algorithmicx}
\usepackage{array}
\usepackage[caption=false,font=normalsize,labelfont=sf,textfont=sf]{subfig}
\usepackage{textcomp}
\usepackage{stfloats}
\usepackage{url}
\usepackage{verbatim}
\usepackage{graphicx}
\usepackage{cite}
\usepackage{makecell}

\usepackage{threeparttable}
\usepackage{diagbox}

\usepackage{color}
\usepackage{wrapfig}
\usepackage{todonotes}
\usepackage{paralist}
\usepackage{units}
\usepackage{booktabs} % For formal tables
\usepackage{latexsym}
\usepackage{multirow}
\usepackage{booktabs}
\usepackage{xspace}
\usepackage[all,pdftex]{xy}
\SelectTips{cm}{}
\usepackage{amsfonts}
\usepackage{textcomp}
\usepackage{xcolor}
\usepackage{mathrsfs}

\usepackage[breakable]{tcolorbox}
\usepackage{balance}

\usepackage{pifont}
\usepackage[noend]{algpseudocode} 

\usepackage{mathtools}
\usepackage{stmaryrd}
\usepackage{color}
\usepackage{colortbl}
\usepackage{stfloats}%for correct floating of two-columns figures
\usepackage{comment}
\usepackage{proof}
\usepackage{fancybox}

\usepackage{hyperref}
\hypersetup{
colorlinks = true, % false: boxed links; true: colored links
linkcolor=blue, % color of internal links
citecolor=blue, % color of links to bibliography
urlcolor=blue, % color of external links
filecolor=black
}
\usepackage{cleveref}

\newcommand{\R}{\mathbb{R}}
\newcommand{\Rpos}{\mathbb{R}_{\ge 0}}

\newcommand{\C}{\ensuremath{\mathcal{C}}\xspace}

\newcommand{\UntilOp}[1]{\mathbin{\mathcal{U}_{#1}}}

\newcommand{\Quan}[2]{\mathtt{rob}\left(#1, #2\right)}

\newcommand{\Defeq}{:=}

\newcommand{\bx}{\ensuremath{\mathbf{x}}\xspace}
\newcommand{\bc}{\ensuremath{\mathbf{c}}\xspace}
\newcommand{\bu}{\ensuremath{\mathbf{u}}\xspace}
\newcommand{\bw}{\ensuremath{\mathbf{w}}\xspace}
\newcommand{\bs}{\ensuremath{\mathbf{s}}\xspace}
\newcommand{\bwp}[1]{\ensuremath{\mathbf{w}_{[#1]}}\xspace}
\newcommand{\bwps}[1]{\ensuremath{\mathbf{w}^*_{[#1]}}\xspace}

\newcommand{\bcL}{t^{\mathtt{L}}_{\bc'}}
\newcommand{\bcU}{t^{\mathtt{U}}_{\bc'}}
\newcommand{\bcp}[1]{\ensuremath{\mathbf{c}_{[#1]}}\xspace}

\newcommand{\taustart}{\ensuremath{\tau_{\mathtt{S}}}\xspace}
\newcommand{\tauend}{\ensuremath{\tau_{\mathtt{E}}}\xspace}
\newcommand{\Bc}{\mathcal{B}_{\bc}}

\newcommand{\drel}{\mathit{d_{rel}}}
\newcommand{\dsafe}{\mathit{d_{safe}}}
\newcommand{\aego}{\mathit{a_{ego}}}
\newcommand{\vset}{\mathit{v_{set}}}
\newcommand{\vego}{\mathit{v_{ego}}}
\newcommand{\afref}{\mathit{AF_{ref}}}
\newcommand{\muset}{\mu_{\mathit{set}}}
\newcommand{\hhref}{\mathit{h_{ref}}}
\newcommand{\hhout}{\mathit{h_{out}}}
\newcommand{\errorset}{\mathit{error_{set}}}

\newcommand{\autorepair}{{\textsc{AutoRepair}}\xspace}

\newcommand{\randoms}{\ensuremath{\mathtt{Random}}\xspace}
\newcommand{\similarity}{\ensuremath{\mathtt{Similar}}\xspace}
\newcommand{\minsat}{\ensuremath{\mathtt{MinSat}}\xspace}

\usepackage{amsmath}
\usepackage{amssymb}
\usepackage{amsthm}

\theoremstyle{plain}
\newtheorem{theorem}{Theorem}[section]

\newtheorem{lemma}[theorem]{Lemma}

\theoremstyle{definition}

\newtheorem{example}[theorem]{Example}
\theoremstyle{remark}

\newcommand{\tbgray}{\cellcolor{gray!25}}

\begin{document}

\title{\autorepair: Automated Repair for AI-Enabled \\Cyber-Physical Systems under Safety-Critical Conditions}

\author{Deyun Lyu, 
Jiayang Song, 
Zhenya Zhang, 
Zhijie Wang, 
Tianyi Zhang, 
Lei Ma and 
Jianjun Zhao
        % <-this % stops a space
\thanks{Deyun Lyu, Zhenya Zhang and Jianjun Zhao are with the Faculty and Graduate School of Information Science and Electrical Engineering, Kyushu University, Fukuoka, Japan (e-mail: zhang@ait.kyushu-u.ac.jp).}
\thanks{Jiayang Song, Zhijie Wang and Lei Ma are with the Department of Electrical and Computer Engineering, University of Alberta, Edmonton, Canada.}
\thanks{Tianyi Zhang is with the Department of Computer Science, Purdue University, West Lafayette, USA.}
}


\maketitle

\begin{abstract}
Cyber-Physical Systems (CPS) have been widely deployed in safety-critical domains such as transportation, power and energy. Recently, there comes an increasing demand in employing deep neural networks (DNNs) in CPS for more intelligent control and decision making in sophisticated industrial safety-critical conditions, giving birth to the class of DNN controllers. However, due to the inherent uncertainty and opaqueness of DNNs, concerns about the safety of DNN-enabled CPS are also surging. In this work, we propose an automated framework named \autorepair that, given a safety requirement, identifies unsafe control behavior in a DNN controller and repairs them through an optimization-based method. Having an unsafe signal of system execution, \autorepair iteratively explores the control decision space and searches for the optimal corrections for the DNN controller in order to satisfy the safety requirements. We conduct a comprehensive evaluation of \autorepair on 6 instances of industry-level DNN-enabled CPS from different safety-critical domains. Evaluation results show that \autorepair successfully repairs critical safety issues in the DNN controllers, and significantly improves the reliability of CPS. 
\end{abstract}


\section{Introduction}
Cyber-Physical Systems (CPS) employ computer technologies to monitor and control physical components  for various real-world tasks. Examples of CPS include adaptive cruise control systems, wind turbine systems, smart grid systems, etc. Typically, a CPS consists of a physical plant and a digital controller---the controller makes control decisions depending on the state of the plant and the external environment, and the plant executes the control decisions to trigger a state transition.

Traditional controllers, such as those based on \emph{proportional integral derivative (PID)}~\cite{johnson2005pid} and \emph{model predictive control (MPC)}~\cite{camacho2013model}, often adopt human-crafted heuristics or linear models to make control decisions. Developing these controllers is costly, which requires huge investment and numerous iterations to ensure their robustness and reliability to handle sophisticated conditions in the real world. Recent studies~\cite{alsalehi2021neural,liu2021recurrent,yaghoubi2020training} have shown that DNN controllers can outperform traditional controllers in terms of robustness and runtime performance in several CPS domains, demonstrating the promise of using DNN controllers as an alternative for traditional controllers. 

Despite an increasing trend of adopting DNN controllers in CPS, a surge of concerns also arise due to the inherent uncertainty and opacity nature of DNNs. As CPS are often deployed in safety-critical domains, improper control decisions may lead to catastrophic system failures, which may further bring intolerable losses and severe social impacts. Most of the existing research efforts have been paid to the testing of CPS (e.g., optimization-based falsification~\cite{Annpureddy-et-al2011, Donze10, falsificationTCAD2018, falsificationCAV2019, falsQBRobCAV2021, constrFalsTCAD2020, gpbConfEstFalsFM21, zhang2022falsifai}), which aims to detect those dangerous cases. However, to enhance the reliability of DNN-enabled CPS and reduce the risks caused by improper control decisions, it is necessary to diagnose and repair the unsafe system behavior after finding them out, which has not received much attention. 

In this paper, we make an early attempt along this direction and propose a framework, \autorepair, for automated repair of DNN-enabled CPS under safety-critical conditions. Given a set of safety requirements, \autorepair first identifies unsafe system behavior by checking the safety satisfaction of system outputs. Then, \autorepair repairs the control decisions through an optimization-based method. Given an unsafe system execution, \autorepair iteratively explores the control decision space and searches for optimal control corrections that enable the system to satisfy the safety requirements. Finally, \autorepair includes those repaired control inputs and outputs into a training dataset and performs robust retraining of the DNN controller. As a result, an enhanced DNN controller is obtained which corrects the unsafe system behavior and improves the reliability of the system under safety-critical conditions.

In summary, we make the following contributions:
\begin{compactitem}[$\bullet$]
\item First, we propose a novel technique for automated repair of DNN-enabled CPS under safety-critical conditions. We formulate the problem into a multi-objective optimization, and employ optimization solvers to search for the optimal corrections for the control signals, which ensure that the system execution after repair satisfies the safety requirement;
\item Then, we implement our proposed technique as a general framework called \autorepair that can handle industrial CPS in \emph{Simulink}, which is the \emph{de facto} standard for CPS modeling environment used in industry;
\item We perform a comprehensive experimental evaluation on 6 instances of industry-level DNN-enabled CPS. The evaluation results demonstrate the effectiveness of our approach.
\end{compactitem}
To the best of our knowledge, this is the first work that aims to repair DNN controllers with respect to system-level safety requirements. Together with the trending demand of adopting DNN in industrial CPS, this work paves the path to further research and real-world safe adoption of DNN in complex industrial CPS.

\section{Preliminaries}
In this section, we give an overview of DNN-enabled CPS (Section~\ref{sec:systemModel}), DNN controllers (Section~\ref{sec:dnnController}), and the system requirements expressed in \emph{Signal Temporal Logic (STL)}
(Section~\ref{sec:safetyReq}) as the preliminaries of our proposed techniques.

\subsection{DNN-Enabled CPS}\label{sec:systemModel}

\begin{figure}[!tb]
\centering
\includegraphics[width=0.7\columnwidth]{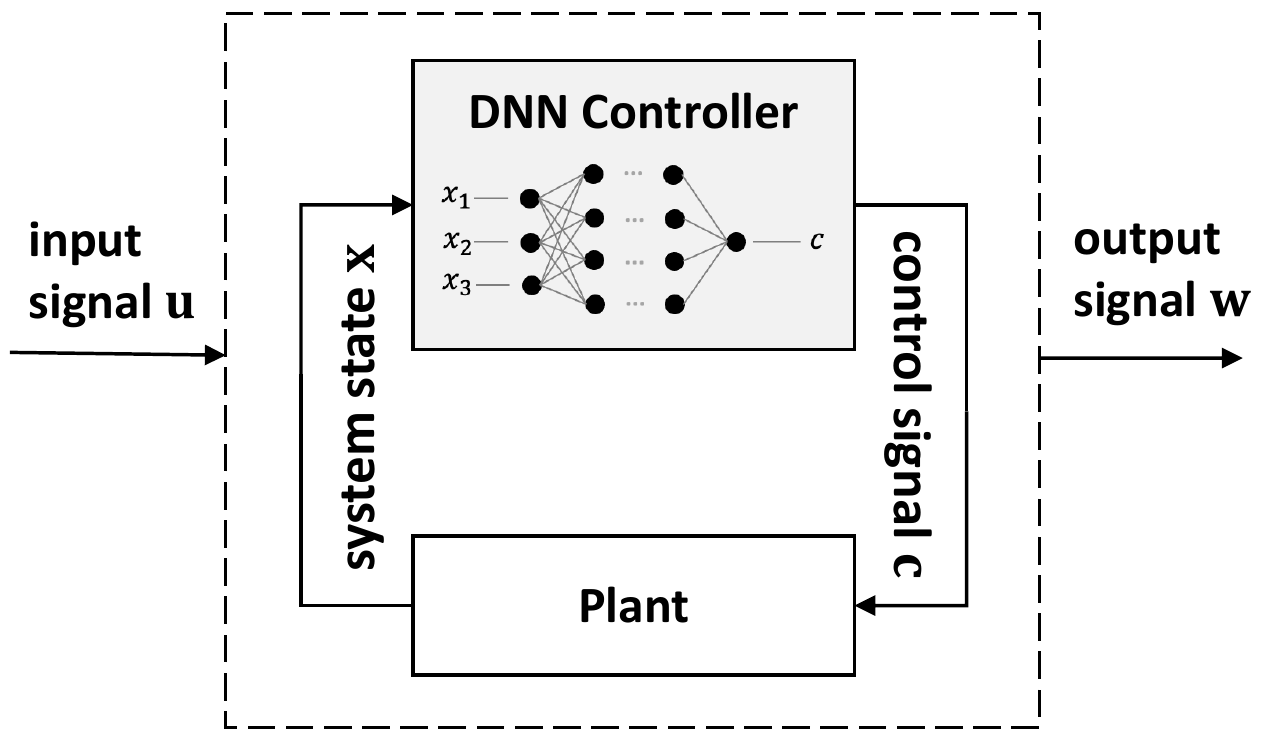}
\caption{The architecture of a typical DNN-enabled CPS}
\label{fig:dnn-cps}
\end{figure}

Fig.~\ref{fig:dnn-cps} shows the typical overall system architecture of DNN-enabled CPS. In particular, the system consists of a {\em plant} and a {\em DNN controller}. The plant is a physical component with continuous state transitions, which are highly non-linear and complex. The DNN controller takes a system state $\bx$ and an external input signal $\bu$ as input. Then, it issues a control signal $\bc$ at each timestamp to control the execution of the plant. This in turn produces an output signal $\bw$ that is observable from the outside. To model the dynamics of signals and system states (i.e., $\bu,\bw,\bx,\bc$), we define them as   \emph{time series}, which maps a time instant $t\in[0,T]$ ($T\in\Rpos$) to a vector $\bs(t)\in \R^d$; Here, $d$ is the dimension of the signal or state, and $[0,T]$ is the \emph{time range} of the signal or state.

Specifically, $\bc$ is one-dimensional, and is bounded by a closed interval $\Bc$ of real numbers, i.e., $\forall t\in[0,T] \,.\, \bc(t)\in \Bc$. 

\subsection{DNN Controllers and Their Training}\label{sec:dnnController} 

Traditional CPS controllers, such as those based on \emph{PID}~\cite{johnson2005pid} and \emph{MPC}~\cite{camacho2013model}, employ human-crafted heuristics or linear models to make control decisions.
Recently, DNN has been increasingly investigated in industrial contexts as an alternative to traditional control algorithms, in handling sophisticated conditions in the real world.
Currently, two approaches are usually adopted to train DNN controllers, namely, \emph{Deep Reinforcement Learning (DRL)} and \emph{Supervised Learning}:
\begin{compactitem}[$\bullet$]
\item DRL allows a trainee to explore in the action space; the trainee will get rewarded if it selects ``good'' actions, and get penalized if it selects ``bad'' actions. Finally the optimal policy will be obtained by the trainee, in the form of a DNN, by iteratively performing ``trial and error''. 
\item In supervised learning, the trainee learns from the historical or experience data collected in the CPS as a training dataset $\{\langle\bx, \bc\rangle\}$, in which each element is a pair of a system state $\bx$ and the corresponding control decision $\bc$. The training phase is to optimize the DNN parameters that maximize the accuracy of predicting $\bc$ given $\bx$. 
\end{compactitem}
In this paper, we consider the DNN controller repair in the CPS development and maintenance contexts under the white-box settings, in order to assist the engineers to identify and repair the potential safety issues at an early stage. In other words, the engineers can access the internals of a DNN controller, so that with more repaired signal data collected either by our techniques or from practical CPS physical environment, it is possible to repair and enhance the safety behavior of CPS, e.g., by retraining or fine-tuning the DNN controller, for continuous delivery. Moreover, we assume that $\bu, \bw, \bx, \bc$ (i.e., the signals between system components and between the system and external environment) are all observable. Furthermore, we assume that the control decisions made by the DNN controller account for the unsafe behavior of CPS.

\subsection{Safety Requirement of CPS}\label{sec:safetyReq}
In this work, we adopt \emph{Signal Temporal Logic (STL)}~\cite{maler2004monitoring} to specify the safety requirements of the DNN-enabled CPS. STL has been a widely-accepted formalism~\cite{ARCHCOMP20Falsification, ARCHCOMP21Falsification} for expressing the users' desired properties. It extends the classic \emph{linear temporal logic (LTL)}~\cite{pnueli1977temporal} in the discrete time setting to continuous time and space domain, and is equipped with a quantitative semantics, called \emph{robustness}, that not only tells \emph{whether} a property is satisfied, but also how much it is satisfied. The quantitative STL semantics has become the technical fundamental of many other safety assurance techniques, such as \emph{runtime monitoring}~\cite{donze2013efficient, deshmukh2017robust, jakvsic2018quantitative} and \emph{falsification}~\cite{Annpureddy-et-al2011, Donze10, falsificationTCAD2018, falsificationCAV2019, falsQBRobCAV2021, constrFalsTCAD2020}. Below we introduce the syntax and robustness of STL.

\smallskip
\noindent\textbf{STL syntax}
In STL, the atomic proposition $\alpha$ is defined as $\alpha \equiv (f(\bw) > 0)$, where $f$ is a function that maps an output signal to a real number. An STL formula $\varphi$ is defined recursively as follows: 
\begin{align*}
    \varphi :\equiv \alpha\mid \bot\mid \neg\varphi \mid \varphi_1\land \varphi_2\mid \Box_I\varphi\mid \Diamond_I\varphi\mid \varphi_1\UntilOp{I}\varphi_2
\end{align*}
where $I$ represents a time interval. Here, $\Box_I\varphi$, $\Diamond_I\varphi$ and $\varphi_1\UntilOp{I}\varphi_2$ are the temporal propositions, where $\Box_I\varphi$ requires 
that \emph{something always happens}, $\Diamond_I\varphi$ requires that \emph{something eventually happens}, and $\varphi_1\UntilOp{I}\varphi_2$ requires that $\varphi_1$ \emph{keeps happening until} $\varphi_2$. 
Other common operators, such as $\lor$, can be derived by the existing operators: $\varphi_1\lor\varphi_2 \equiv \neg(\neg\varphi_1 \land \neg\varphi_2)$.

\smallskip
\noindent\textbf{STL robustness}
Given the system output $\bw$ and an STL formula $\varphi$, the robustness $\Quan{\bw}{\varphi}$ returns a real number that indicates \emph{how robustly} $\bw$ satisfies $\varphi$. The definition of $\Quan{\bw}{\varphi}$ is described as follows, by induction on the structure of the STL formula:
\begin{align*}
    &\Quan{\bw}{\alpha} \Defeq f(\bw) \qquad \Quan{\bw}{\bot} \Defeq -\infty \\
    &\Quan{\bw}{\neg\varphi} \Defeq -\Quan{\bw}{\varphi} \\
    &\Quan{\bw}{\varphi_1\land \varphi_2} \Defeq \min\left(\Quan{\bw}{\varphi_1}, \Quan{\bw}{\varphi_2}\right) \\
    &\Quan{\bw}{\Box_I\varphi} \Defeq \inf_{t\in I}(\Quan{\bw^t}{\varphi}) \\
    %&\Quan{\bw}{\Diamond_I\varphi} \Defeq \sup_{t\in I}(\Quan{\bw^t}{\varphi}) \\
    &\Quan{\bw}{\varphi_1\UntilOp{I}\varphi_2} \Defeq \sup_{t\in I} \min\left(
    \begin{array}{l}
      \Quan{\bw}{\varphi_2} ,   \\
    \inf_{t'\in [0, t)}\Quan{\bw^{t'}}{\varphi_1}
    \end{array}
     \right)
\end{align*}
Here $\bw^t$ denotes a signal obtained by \emph{shifting} the starting point of $\bw$ for $t$ time units. Intuitively, the larger $\Quan{\bw}{\varphi}$ is, the more robustly $\bw$ satisfies $\varphi$; once $\Quan{\bw}{\varphi}$ is negative, then it indicates that $\bw$ violates $\varphi$.

\section{The \autorepair Framework}\label{sec:framework}

\begin{figure*}[!tb]
\centering
% \begin{center}
    
\includegraphics[width=\textwidth]{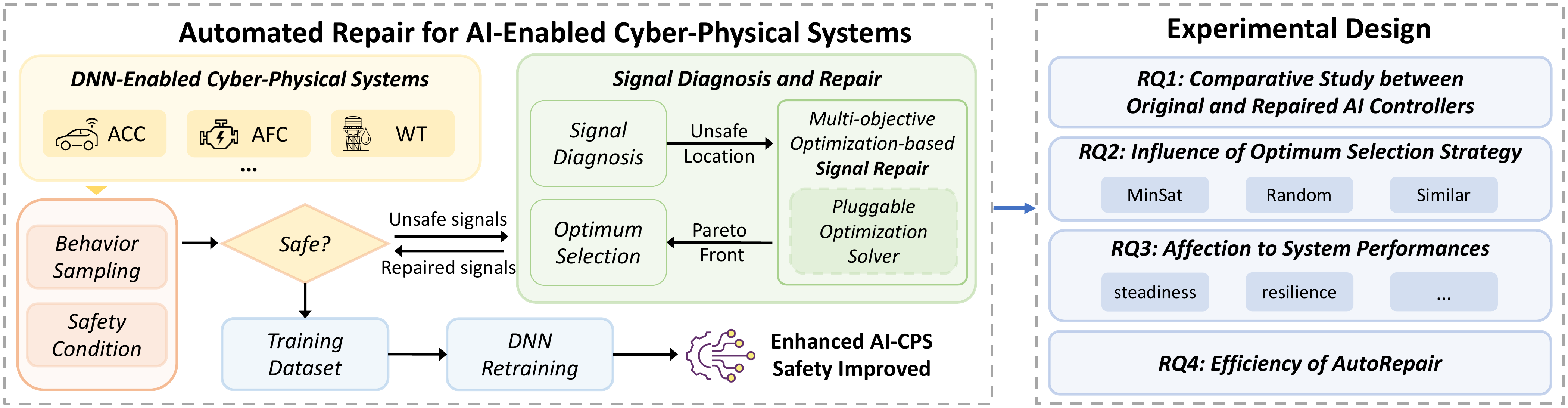}
\caption{The workflow and study design of our proposed framework \autorepair for automated repair of DNN-enabled CPS.}
\label{fig:workflow}
\end{figure*}
In this paper, we propose a framework, named \autorepair, to enhance DNN controllers for system safety. Specifically, \autorepair 
\begin{inparaenum}[(i)]
\item identifies unsafe output signals in a system execution,
\item repairs them by searching for a control signal patch to ensure that the system satisfies safety requirements during the given execution, and
\item repairs the DNN controller with the safe execution trace data.
\end{inparaenum}
We first illustrate how to repair an unsafe system execution in Example~\ref{eg:acc}, and then introduce the workflow of \autorepair.

\begin{figure}[!tb]
\centering
\includegraphics[width=0.85\columnwidth]{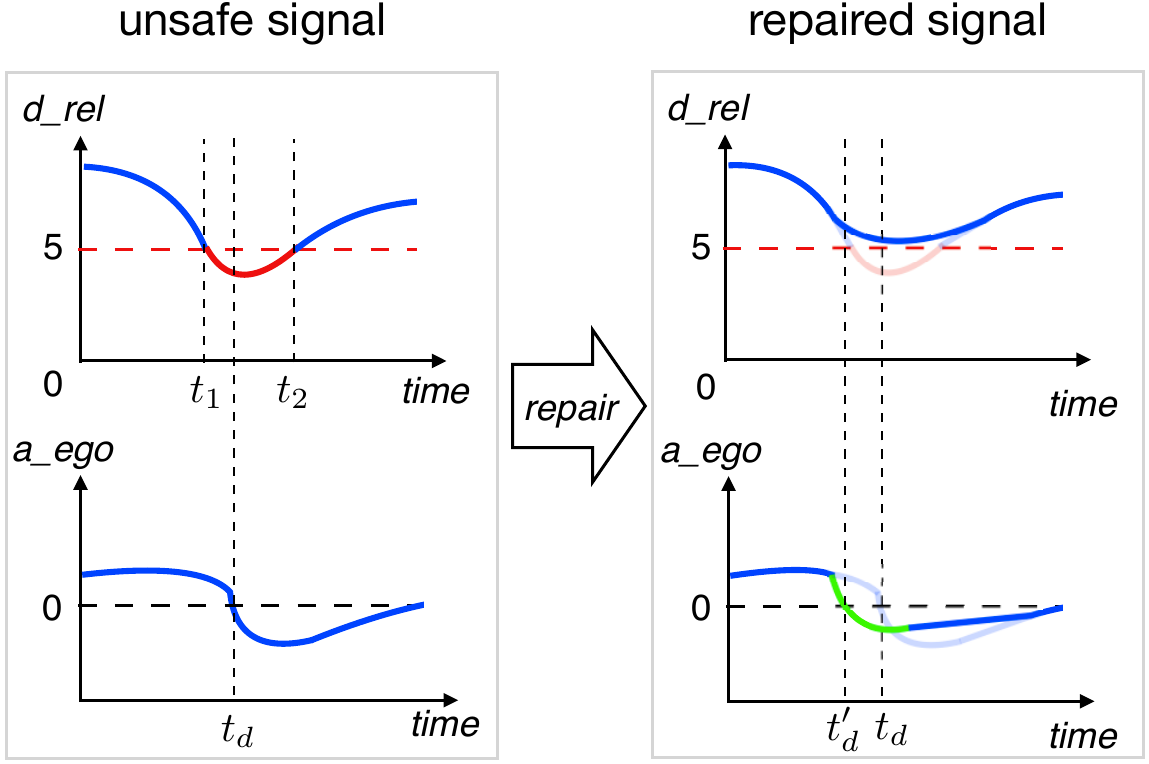}
\caption{An illustrative example of signal repair for an adaptive cruise control (ACC) system.}
\label{fig:ACCEg}
\end{figure}

\begin{example}\label{eg:acc}
Consider the \emph{adaptive cruise control (ACC)} scenario, in which an ego car tries to follow a lead car within a safe distance. We treat the ego car as the DNN-enabled system under consideration, and the lead car as the external environment. 
The controller of the ego car decides its acceleration or deceleration $\aego$ based on the sensor data, including the states of the ego car and the lead car. The system requirement $\varphi$ in this scenario can be expressed as follows, 
\begin{align*}
    \varphi \equiv \Box_{[0, T]}(\drel > 5)
\end{align*}
which requires that, the relative distance $\drel$ between the two cars should always be larger than a safe distance of 5.

The left part of Fig.~\ref{fig:ACCEg} shows an example of system execution that violates the safety requirement. During $[t_1, t_2]$, $\drel$ is smaller than 5. This is because, $t_d$ (i.e., the time when the ego car starts to decelerate) is too late to avoid a close relative distance. The right part of Fig.~\ref{fig:ACCEg} shows an example of signal repair, in which a \emph{control signal patch} (the green segment) replaces the original control signal. The  signal patch brings the deceleration moment $t_d$ forward to $t'_d$. As a result, the ego car decelerates timely and avoids violating the safety requirement. Now, with the control signal patch, we can rerun the system to obtain a new execution trace that avoids the unsafe behavior. This new execution trace will be used to repair the DNN controller to enhance the system reliability.
\end{example}

This example explains the main task of \autorepair, namely, for an unsafe system execution, it requires to synthesize a control signal patch, which enables the system execution to satisfy the safety requirement. Before the technical details, we first overview the high-level workflow of \autorepair.

\medskip
\noindent{\bf Workflow.}
Given a CPS with a DNN controller, \autorepair first runs the system and performs behavior sampling to obtain execution traces, each of which includes a system input signal $\bu$, a system output signal $\bw$, a control decision signal $\bc$, and a system state signal $\bx$. Then, given a safety requirement~$\varphi$, \autorepair checks whether each system output signal $\bw$ in the execution trace satisfies the safety requirement. If unsafe signals are found, the signal diagnosis and repair procedure will be triggered to repair unsafe signals and produce a new safety-assured execution trace. This procedure will be detailed later in Section~\ref{sec:dataGeneration}.

With the safety-assured execution trace, we can repair the original DNN controller that issued improper control signals and led to unsafe system behavior. To be general, as an early step for DNN controller repair, the DNN repair module in \autorepair is designed to be extensible (to diverse repairing methods). The basic mode is to retrain the original DNN controller with the new safety-assured execution trace to enhance its performance, while keep the remaining training configurations unchanged. Although this mode has the lowest cost, it sometimes produces a less robust DNN controller since the model architecture may not be optimal.  In addition to the basic mode, other options include \emph{Neural Architecture Search (NAS)}~\cite{elsken2019neural}, which can search over different DNN configurations and identify the best model with the new execution trace. In this work, we experimentally evaluate the performance of the repaired DNN controller obtained by retraining, and we leave the investigation on other options as future work.

As signal diagnosis and repair are the major technical contributions of the \autorepair framework, we elaborate on their details in the next section. 

\section{Signal Diagnosis and Repair}\label{sec:dataGeneration}
In this section, we introduce the details of the key component of \autorepair, i.e., signal diagnosis and repair, which aims to repair the control signals that lead to unsafe system output in a system execution.

\begin{algorithm}[!tb]
\footnotesize
\caption{The optimization-based signal repair}
\label{algo:traceRepair}
\begin{algorithmic}[1]
\Require The original system with a DNN controller $\C$, a safety requirement $\varphi$, and an execution trace including input signal $\bu$, output signal $\bw$, control signal $\bc$, and state signal $\bx$
\Statex
\Procedure{DiagAndRepair}{$\bu$, $\bw, \bc, \bx$}
\If{$\Quan{\bw}{\varphi} <  0$}\label{line:checksat}\vspace{0.5em}\Comment{$\bw$ is unsafe}
\State $\left[
\begin{array}{l}
    \vspace{0.5em}
    \taustart\gets 
    \text{when the violation episode starts} \\
    \tauend \gets
    \text{when the violation episode ends}
\end{array}
\right]
$\label{line:diagnosis}
\vspace{0.5em}
\Comment{diagnosis}
\State $\bc' \gets \arg
\left[ 
\begin{array}{rl}
   \underset{\bc'}{\textrm{max}}  &\Quan{\bwps{0,\tauend}}{\varphi}\\\vspace{0.5em}
      \underset{\bc'}{\textrm{min}}  & \mathsf{dist}\left(\bc', \bcp{\bcL, \bcU}\right) \\\vspace{0.5em}
     \textrm{s.t.}  & \bcL\in [0, \taustart] \\\vspace{0.5em}
    & \bcU\in (\bcL, \tauend] \\
    & \forall t\in\left[\bcL,\bcU\right].\bc'(t)\in \Bc
\end{array}
\right]$ \label{line:repair}
\vspace{0.3em}
\Comment{repair}
\State $\bw, \bc, \bx \gets$ \textsc{SysExe}\big($\bu, \bc^*$\big) \text{with }$\bc^* = \langle\C, \bc'\rangle$ \label{line:updatewcx} 
\vspace{3pt}
\Comment{system rerun}
\State \Call{DiagAndRepair}{$\bu, \bw, \bc, \bx$}\label{line:recursiveCall} 
\Comment{recursive call}
\EndIf
\EndProcedure
\State add the repaired {$\langle\bx,\bc\rangle$} to the retraining dataset \label{line:return}
\end{algorithmic}
\end{algorithm}

\subsection{Algorithm Overview}
Alg.~\ref{algo:traceRepair} describes the signal diagnosis and repair algorithm. It takes as input a CPS with a DNN controller $\C$, an STL safety requirement $\varphi$, and an execution trace including input signal $\bu$, output signal $\bw$, control signal $\bc$, and state signal $\bx$. Note that all these signals are defined as time series to model their dynamics over a time range $[0,T]$, as defined in Section~\ref{sec:systemModel}.
This procedure starts with checking whether the output signal $\bw$ satisfies the safety requirement $\varphi$ (Line~\ref{line:checksat}). If $\bw$ is safe, then there is no need to repair it. Otherwise, the algorithm first localizes the starting and ending time instants of the unsafe output signal (Line~\ref{line:diagnosis}). Then, it searches for an optimal control signal patch $\bc'$ that corrects the system output signal to satisfy the safety requirement (Line~\ref{line:repair}). Formally, $\bc'$ is a partial signal defined in the time interval $[\bcL, \bcU]$, where $\bcL$ is the \emph{lower time bound} at which the control must start being rectified to avoid the unsafe system behavior, and  $\bcU$ is the \emph{upper time bound} at which the control rectification can end. The details of deciding $\bcL$ and $\bcU$ will be given in Section~\ref{sec:signalDiagnosis}. 

With the control signal patch $\bc'$, we employ an alternate control between the original controller $\C$ and the patch $\bc'$ to re-execute the system, described as below:
\begin{compactitem}[$\bullet$]
\item in $[0,\bcL]$, the control is given by the original controller;
\item in $[\bcL, \bcU]$, the control is switched to $\bc'$;
\item in $[\bcU, T]$, the control is switched back to the original controller, until the end of the system execution.
\end{compactitem}

Using this control switch, we can obtain new $\bw$, $\bc$ and $\bx$ by re-executing the system (Line~\ref{line:updatewcx}). This procedure is invoked recursively with the new $\bw, \bc$ and $\bx$ to diagnose and repair new unsafe behavior, until the entire system execution is safe. Finally, this new execution is returned to the \autorepair framework and fed as input into the DNN repair module to enhance the DNN controller (Line~\ref{line:return}).

\subsection{Signal Diagnosis}\label{sec:signalDiagnosis}
A naive selection of the time range $[\bcL, \bcU]$, in which we search for the control signal patch $\bc'$ to repair the unsafe system execution, is the entire time range $[0,T]$. 
However, the search in that case will be too expensive, due to the large search space. 
In this section, we leverage the recent progress of STL \emph{signal diagnosis}~\cite{zhang2022online, bartocci2018localizing} to reduce the search space. 

Signal diagnosis is a technique that can
localize the \emph{violation episode} in a signal. Intuitively, a violation episode is a segment of a signal that could be considered as the \emph{cause} of the violation of the whole signal to a given safety property. In general, signal diagnosis is achieved by recursively collecting the time instants at which the evaluations of the atomic proposition of the safety property lead to the violation of the given STL formula, based on the semantics of different STL operators. We illustrate this process using an example in Example~\ref{eg:diagnosis}. For more details, we refer the readers to~\cite{zhang2022online}.

\begin{example}\label{eg:diagnosis}
In Example~\ref{eg:acc}, the segment in $[t_1, t_2]$ could be considered as the cause of the violation, because $\drel$ keeps being less than 5 in this segment, which violates the atomic proposition $\drel > 5$ of $\varphi$. As a result, the violation of the atomic proposition leads to the violation of $\varphi$, and therefore, this segment will be diagnosed as a violation episode.
\end{example}

In Line~\ref{line:diagnosis} of Alg.~\ref{algo:traceRepair}, $\taustart$ is set as the timestamp when the violation episode starts, i.e., when the system output starts to violate the safety requirement; and $\tauend$ is set as the timestamp when the violation episode ends, i.e., when the system resumes back to a safe status.  The correctness of using signal diagnosis to derive the time constraints of controller repair is based on the fact that the unsafe status caused by improper control decisions only propagate backward in the time domain, and therefore, only the time interval that precedes or matches the safety violation episode of the output signal needs to be considered for repair. See Lemma~\ref{prop:bound} for an explanation.

\begin{lemma}\label{prop:bound}
Let $\bc'$ be a control signal patch defined in the interval $[\bcL, \bcU]$, and $\taustart$ and $\tauend$ be respectively the starting and the ending point of the violation episode.
Then, the lower time bound $\bcL$ of  $\bc'$ should be not later than $\taustart$; and the upper time bound $\bcU$ of $\bc'$ should  be not later than $\tauend$.
\end{lemma}

The proof of Lemma~\ref{prop:bound} is based on the definition of STL signal diagnosis, and the fact that safety violations only propagate backward in the time range. 

If $\bcL > \taustart$, there will be an unsafe segment in $[\taustart, \bcL]$ that is not repaired by the control signal patch $\bc'$. Therefore, the lower time bound $\bcL$ of $\bc'$ should not be later than $\taustart$. 
Similarly, we can derive that the upper time bound $\bcU$ of $\bc'$ should not be later than $\tauend$. \qed

\begin{example}
In Example~\ref{eg:acc}, $\taustart$ and $\tauend$ are respectively computed as $\taustart = t_1$ and $\tauend = t_2$. Therefore, when we search for the control signal patch $\bc'$, its lower time bound $\bcL$ should be not later than $t_1$, and its upper time bound $\bcU$ should be not later than $t_2$.

Intuitively, this is correct, because if $\bcL$ is later than $t_1$, then the safety violated interval $[\taustart, \bcL]$ cannot be repaired by $\bc'$; if $\bcU$ is later than $\tauend$, then in $[\tauend, \bcU]$ there is no safety violation.
\end{example}

\subsection{Multi-Objective Optimization-Based Repair}\label{sec:signalRepair}
There remains the problem of how to construct the control signal patch $\bc'$, given the time constraint $[\bcL, \bcU]$. 
%and space (see \S{}\ref{sec:systemModel}). 
The desired properties of $\bc'$ involve the following two aspects:
\begin{compactitem}[$\bullet$]
\item First, by substituting the original control signal with $\bc'$, the safety violation in $[\taustart, \tauend]$ should be resolved;
\item Second, the change between the original control signal $\bc$ and the control signal patch $\bc'$ should be minimal, in order to minimize the performance losses in other aspects than safety and also avoid producing severe outlier control signals that harm the quality of training data.
\end{compactitem}

In this paper, we search for such a control signal patch $\bc'$ using \emph{multi-objective optimization}. In general, multi-objective optimization consists of multiple objective functions; these functions may conflict with each other, i.e., optimizing one function may hinder another function. As a result, the outcome of multi-objective optimization is a set of optima, called \emph{Pareto front}, rather than a single optimum. Intuitively, a Pareto front involves all the incomparable optimal solutions---it is often not the case that one solution is better than another one in terms of all the objective functions.

We formulate our problem of searching for $\bc'$ as multi-objective optimization, as in Line~\ref{line:repair} of Alg.~\ref{algo:traceRepair}: \begin{compactitem}[$\bullet$]
\item The first objective function is to maximize the satisfaction of the repaired output signal $\bwps{0,\tauend}$ in $[0, \tauend]$ to the safety requirement. Initially, $\bwps{0,\tauend} = \bwp{0,\tauend}$ and so $\Quan{\bwps{0,\tauend}}{\varphi}$ is negative. The goal of this objective function is to find a  $\bc'$ such that the satisfaction  $\Quan{\bwps{0,\tauend}}{\varphi}$ of the repaired signal $\bwps{0,\tauend}$ is positive, and so $\bwps{0,\tauend}$ satisfies the safety requirement $\varphi$.
\item The second objective function is to minimize the change of $\bc'$ compared to the original control signal in $[\bcL, \bcU]$, in terms of a specific distance measure.
\item The constraints include the time constraint in Section~\ref{sec:signalDiagnosis} and the space constraint in Section~\ref{sec:systemModel}.
\end{compactitem}
An issue arises in handling the infinite search space for $\bc'$ due to its continuity in time and space. Following the convention of CPS community~\cite{ARCHCOMP20Falsification, ARCHCOMP21Falsification, ARCHCOMP22Falsification}, we adopt a \emph{parameterized representation} to denote $\bc'$, so the continuous signal $\bc'$ can be identified by a finite number of variables. Specifically, we select \emph{piecewise linear} signal, i.e., given a hyperparameter $k$, $\bc'$ is linear in each interval $\left[\bcL + \frac{i(\bcU-\bcL)}{k}\,,\, \bcL + \frac{(i + 1)(\bcU-\bcL)}{k}\right]$ ($i\in\{1, \ldots, k-1\}$). In that case, $\bc'$ is identified by $\bcL, \bcU$, and the $k$ variables that characterize the shape of $\bc'$. %\todo{interp, nolinear}

An efficient optimization solver is needed to solve the optimization problem formulated in Line~\ref{line:repair}. In our context, the evaluation of the objective $\Quan{\bwps{0,\tauend}}{\varphi}$ requires system execution of CPS, whose dynamics is often hard to model. To that end, we select \emph{NSGA-II}~\cite{deb2002fast}, a genetic algorithm (GA)-based multi-objective optimization algorithm, as our solver, since NSGA-II does not require the awareness of the internal dynamics of the objective function. It computes a Pareto front by iteratively performing \emph{mutation} and \emph{crossover}, as other genetic algorithms do.

The Pareto front returned by the optimization solver contains a set of optimum solutions; having such a Pareto front, we need to further perform an \emph{optimum selection} to select the solution used for signal repair. In \autorepair, we make this selection process configurable, and here we provide three possible strategies, as follows. In RQ2 of Section~\ref{sec:experimentResults}, we experimentally compare their performances.
\begin{compactitem}
    \item \minsat selects the optimum which triggers a minimum satisfaction to the safety requirement, to accommodate the other objective function;
    \item \randoms refers to a na\"{i}ve selection strategy that randomly selects a solution from the Pareto front, as long as it satisfies the safety requirement;
    \item \similarity selects the solution with the minimal change w.r.t. the original control decisions, under the premise of satisfying the safety requirement.
\end{compactitem}

After the error-triggering signals are repaired, the next step is to leverage such signal data for the DNN controller repair, as described in Section~\ref{sec:framework}. In general, \autorepair can be easily integrated into the development and maintenance of DNN-enabled CPS for DNN repair and enhancement, in order to improve their safety.

\section{Experiment Design and Setup}\label{sec:experimentSetup}

 In this section, we introduce the design and setup of the experiments used to evaluate the usefulness and performance of \autorepair. All the source code and benchmarks are publicly available: {\url{https://github.com/lyudeyun/AutoRepair}}.

\begin{table*}[!tb]
% \footnotesize
\centering
\caption{Subject CPS adopted to evaluate \autorepair. The column \textbf{Blocks} reports the number of the component blocks in a system, in order to measure the complexity of the model.}
% \end{center}
\label{tab:bench}
% line height
\renewcommand{\arraystretch}{1.2}
\resizebox{\textwidth}{!}{
\begin{tabular}{llllc}
\toprule
\textbf{Subject CPS} & \textbf{Description} & \textbf{Input signals} & \textbf{Control signals} & \textbf{Blocks} \\\hline
  Adaptive Cruise Control   &     Maintain a safe distance from a lead car        &   Lead car acceleration    &      Ego car acceleration          &  297         \\
 Abstract Fuel Control     &   Maintain the reference air-to-fuel ratio          &   Pedal angle \& engine speed    &      Fuel command       & 281           \\
 Water Tank     &     Keep the water level at a reference value        &    Reference water level   &    Water flowrate    &            919       \\\bottomrule
\end{tabular}
}
\end{table*}

\subsection{Research Questions}
We design our experiments in order to investigate the following research questions: 
\begin{compactitem}[$\bullet$]
\item \medskip\textbf{RQ1: Does \autorepair effectively improve the safety of DNN-enabled CPS?} 

In this RQ, we aim to evaluate whether the subject CPS with the DNN controller repaired by \autorepair can indeed improve the safety of the system. To answer this question, we first create the retraining dataset by sampling the execution traces of these systems, and apply \autorepair to obtain the repaired systems with retrained DNN controllers. For the testing purpose, we also create a test set for each system. Then, we execute the original systems and the repaired systems, by feeding them the input signals from the retraining dataset and the test set. We compare the safety rates of these systems, which are measured as the ratio of the number of safe executions over the number of all the executions in the retraining dataset and the test set.

\item \medskip\textbf{RQ2: How does the optimum selection strategy influence the effectiveness of \autorepair?}

A key phase in \autorepair is the signal diagnosis and repair procedure introduced in Section~\ref{sec:dataGeneration}. Given an unsafe execution trace, we search for an optimal control signal patch which ensures the system safety. However, since the output given by the optimization solver is a Pareto front, we need to select a specific optimum and apply it as the control signal patch to repair the system. In Section~\ref{sec:signalRepair}, we have introduced three optimum selection strategies, namely, \minsat, \randoms, \similarity. In this RQ, we compare the safety rates of the systems under different optimum selection strategies to study the impact of these strategies.

\medskip
\item \textbf{RQ3: Does \autorepair affect the performance of DNN-enabled CPS in terms of other metrics than safety?}

Although \autorepair aims to improve the system safety, a concern arises that it may diminish the system performances in other aspects, such as the efficiency of the system execution. In this RQ, we investigate the influence of \autorepair on the performance of the DNN-enabled CPS w.r.t. various system performance metrics. Specifically, we compare the performance of the repaired systems, by \autorepair with the default optimum selection strategy \minsat, with the original systems in terms of these metrics. Note that these system performance metrics are system-specific, since different systems are constrained by different requirements in their development. A detailed introduction to these metrics is given in Section~\ref{sec:benchmark}. 

\medskip
\item \textbf{RQ4: How is the efficiency of \autorepair?}

In addition to the effectiveness of \autorepair, the efficiency of \autorepair is another important factor for its real-world application.
The most time consuming part in \autorepair is the procedure of signal repair for each unsafe system execution. Since this procedure relies on the multi-objective optimization algorithm NSGA-II, the time cost is dependent on the configuration of NSGA-II. Moreover, different systems have different execution time, and this also affects the time cost of applying \autorepair. In this RQ, we investigate this aspect of \autorepair, and report how efficient \autorepair is for repairing the unsafe behavior of DNN-enabled CPS.
\end{compactitem}

\subsection{Benchmarks}\label{sec:benchmark}

 We selected DNN-enabled CPS from three industrial domains as the subject systems for our evaluation, namely, \emph{Adaptive Cruise Control} (ACC)~\cite{ACCMathworks}, \emph{Abstract Fuel Control} (AFC)~\cite{jin2014powertrain}, and \emph{Water Tank} (WT)~\cite{WTMathworks}. These systems span over different safety-critical industrial domains, such as autonomous driving and powertrain control. Table~\ref{tab:bench} summarizes these systems with a short description of their functionalities. All these systems are built in Simulink, the \emph{de facto} CPS modeling standard in industry. The rest of this section contains the detailed information of each subject CPS, which involves their functionality, safety requirements and other performance metrics.
 
\medskip
\noindent\textbf{Adaptive Cruise Control (ACC)}. As an important driving assistance function, ACC has been popularized in the automotive field after decades of development. The ACC system mentioned in this work is originally from MathWorks~\cite{ACCMathworks}, and it aims to maintain the safety distance of an ego car from a lead car by adjusting the acceleration of ego car. When the relative distance is larger than the safe distance, the speed of ego car will instead maintain at the driver-set velocity. To this end, the whole system takes the acceleration of the lead car as input and outputs the velocity and the moving distance of the lead car and the ego car. The safety requirement of ACC is formulated as follows, saying that during the simulation, the relative distance, $\drel$, between the two cars should always be larger than a safety distance, defined by the sum of a constant $\dsafe$ and the braking distance of the ego car. Meanwhile, the speed of the ego car $\vego$ should be lower than $\vset$. Here, we set $\dsafe$ and $\vset$ as 10 and 30, respectively. 
\begin{align*}
S_{\mathrm{ACC}} \equiv \Box_{[0,50]} (\drel \geq \dsafe \wedge \vego \leq \vset)
\end{align*}
The performance metrics of ACC are described as follows. 
\begin{compactitem}
\item \emph{space} refers to the longitudinal distance traveled by the ego car. The greater the \emph{space} is, the higher the driving efficiency is;
\item \emph{speed} denotes the average deviation between the actual speed of the ego car and the expected speed when the car travels in a safe status. A smaller \emph{speed} means higher cruise efficiency;
\item \emph{steadiness} aims to measure the system stability, which is reflected by the proportion of the time period during which the system satisfies the safety requirement over the total time interval $[0,50]$; 
\item \emph{resilience} aims to assess whether a controller acts rapidly to return to a steady state, when the system is not steady; 
\item \emph{comfort} indicates how comfortable the ego car is, by measuring the maximum changing rate of the acceleration during the system execution.
\end{compactitem}
 
 \medskip
\noindent\textbf{Abstract Fuel Control (AFC)}. AFC is a complex air-fuel control system released by Toyota~\cite{jin2014powertrain}. The whole system takes two input signals from the outside environment, namely, pedal angle and engine speed, and outputs $\mu = \frac{|AF - \afref|}{\afref}$, which is the deviation of the air-to-fuel ratio AF from a reference value $\afref$. By changing the system inputs, the fuel controller should adjust the intake gas rate to the cylinder to maintain the optimal air-to-fuel ratio. The goal of this system is to control the deviation $\mu$ such that it is within a predefined threshold $\muset = 0.15$ during the time interval $[0,30]$. 
\begin{align*}
S_{\mathrm{AFC}} \equiv \Box_{[0,30]} \left(\frac{|AF - \afref|}{\afref} \leq \muset \right)
\end{align*}
The performance metrics of AFC are described as follows.
\begin{compactitem}
\item \emph{error} indicates the absolute value of the actual deviation rate $\mu$. A small \emph{error} indicates better performance, and ideally, \emph{error} should be 0; 
\item \emph{steadiness} is characterized by the proportion of the time period during which the system satisfies a safety requirement over the total time interval $[0,30]$.
\end{compactitem}

\medskip
\noindent\textbf{Water Tank (WT)}. As a container for controlling the inflow and outflow of water, WT has been applied in the domains such as chemical industry. This system is collected from MathWorks~\cite{WTMathworks}. Water can inflow or outflow the water tank, until the height of water in the tank reaches the reference value. After a certain time, the height of water should be the same as the reference height. The system takes the reference water level $\hhref$ as input signal and outputs the actual water level $\hhout$ in real time. The absolute deviation $error$ between $\hhref$ and $\hhout$ should always be less than a setting value $\errorset$ within the time intervals $[4,5]$, $[9,10]$ and $[14,15]$. Here $\errorset$ is 0.86.
\begin{align*}
    S_\mathrm{WT} &\equiv \Box_{\mathrm{I}} (|error| \leq \errorset) \\
    & \text{where }\mathrm{I} = {[4,5] \cup [9,10] \cup [14,15]}
\end{align*}
The performance metrics of WT are formulated as follows, 
\begin{compactitem}
    \item \emph{error} indicates the absolute error value between the reference value and the real value. For WT, this is indicated by the absolute error value between $\hhref$ and $\hhout$. A smaller \emph{error} represents a higher control accuracy;
    \item \emph{steadiness} is  characterized by the proportion of the time period during which the system satisfies a safety requirement, over the time intervals $[4,5]$, $[9,10]$ and $[14,15]$.
\end{compactitem}

\medskip
\noindent\textbf{Neural network controllers}. 
To demonstrate the effectiveness of \autorepair, for each subject CPS, we trained 2 neural network controllers with different structures, and thus, we have 6 instances in total as the systems under evaluation. In Table~\ref{tab:details}, we make a detailed introduction to the original controllers deployed on these systems, including their types, structures and training algorithms. Taking ACC\#2 as an example, the original DNN controller constructed by \emph{Feed-Forward Neural Network (FFNN)} consists of 3 hidden layers, and each hidden layer has 30 neurons. This controller is trained by the built-in BFG algorithm of \emph{Deep Learning Toolbox} in MATLAB. 
 
\medskip
\noindent\textbf{Retraining set and test set}. For each instance of the subject systems, we randomly sample 10000 execution traces as the dataset for retraining, and 1000 traces as the test set. We list the safety rates (SR) of the systems with original controllers in the retraining dataset in Table~\ref{tab:details}, which uses $\mathit{\frac{\# safe\ trials}{\# total\ trials}}$ to represent how safe a system is. As illustrated in Table~\ref{tab:details}, the safety rates of the 6 instances to be repaired range from 82.0$\%$ to 98.1$\%$.

\begin{table}[!tb]
\centering
\caption{The details of six subject CPS instances}
\label{tab:details}
% line height
\renewcommand{\arraystretch}{1.2}
\begin{threeparttable}
\begin{tabular}{ccc}
\toprule
\textbf{Instance}    & \textbf{Original DNN Controller}    & \textbf{SR}                          \\ \hline
\multirow{3}{*}{ACC\#1} & Type: FFNN                  & \multirow{3}{*}{9682/10000} \\
                        & Structure: {[}10, 10, 10{]} &                            \\
                        & Training Algorithm: LMBP\tnote{*}     &                            \\ \hline
\multirow{3}{*}{ACC\#2} & Type: FFNN                  & \multirow{3}{*}{8200/10000} \\
                        & Structure: {[}30, 30, 30{]} &                            \\
                        & Training Algorithm: BFG\tnote{**}     &                            \\ \hline
\multirow{3}{*}{AFC\#1} & Type: FFNN                  & \multirow{3}{*}{8287/10000} \\
                        & Structure: {[}15, 15, 15{]} &                            \\
                        & Training Algorithm: LMBP     &                            \\ \hline
\multirow{3}{*}{AFC\#2} & Type: FFNN                  & \multirow{3}{*}{8661/10000} \\
                        & Structure: {[}30, 30, 30{]} &                            \\
                        & Training Algorithm: LMBP     &                            \\ \hline
\multirow{3}{*}{WT\#1}  & Type: FFNN                  & \multirow{3}{*}{8404/10000} \\
                        & Structure: {[}5, 5, 5{]} &                            \\
                        & Training Algorithm: BFG     &                            \\ \hline
\multirow{3}{*}{WT\#2}  & Type: FFNN                  & \multirow{3}{*}{9808/10000} \\
                        & Structure: {[}15, 15, 15{]} &                            \\
                        & Training Algorithm: BFG     &                            \\ \bottomrule
\end{tabular}
\begin{tablenotes}
        \footnotesize
        \item[*] LMBP: \emph{Levenberg–Marquardt Backpropagation (LMBP) Algorithm}~\cite{lv2018trainlm}
        \item[**] BFG: \emph{BFGS quasi-Newton (BFG) Algorithm}~\cite{gill2019trainbfg}
\end{tablenotes}
\end{threeparttable}
\end{table}

\subsection{Software and Hardware Dependencies}
All the subject CPS are implemented in Simulink with dependencies on MATLAB toolboxes {\em Model Predictive Control}, {\em Control System}, and {\em Deep Learning}.   The experiments were conducted on AWS cloud instances of type c4.2xlarge (2.9 GHz Intel Xeon E5-2666 v3, 15G RAM). For DNN controller retraining, we employ the features of the \emph{Deep Learning Toolbox} of MATLAB, and run it on a server with a 3.3GHz Intel i9-10940X CPU, 64G RAM, and two NVIDIA RTX A6000 GPUs.
The evaluation took around 480~hours (20~days~$*$~24~hours) on 20 AWS instances in total.

\begin{table}[!tb]
\centering
\caption{Parameters of NSGA-II Algorithm}
\label{tab:nsga-ii}
\renewcommand{\arraystretch}{1.2}
\begin{tabular}{lc}
\toprule
\textbf{Parameter of NSGA-II}       & \textbf{Value} \\ \hline
Population Size             & 40    \\ 
Maximum Generation          & 10    \\ 
Number of Objective         & 3     \\ 
Number of Decision Variable & 6     \\ \bottomrule
\end{tabular}
\end{table}
To conduct multi-objective optimization-based repair  in Section~\ref{sec:signalRepair}, we select NSGA-II as our solver. The main parameters of NSGA-II employed in our evaluation are listed in Table~\ref{tab:nsga-ii}. Here we set the population size and the maximum generation of NSGA-II as 40 and 10, respectively. The number of objective functions is 3, with the second one (regarding the distance between the original control signal and the control patch signal) in Section~\ref{sec:signalRepair} instantiated to a spatial constraint and a temporal constraint. The number of decision variables is the dimension of a single optimal solution, namely 6, which consists of 2 timestamps $\bcL$ and $\bcU$, and 4 interpolation points used to generate the control signal patch $\bc'$.

\begin{table*}[!tb]
\centering
\caption{The safety rate of the original systems, and the systems after repair under three optimum selection strategies. The best performer for each system instance is highlighted. (\emph{Train}: the dataset for retraining. \emph{Test}: the test set.)}
\label{tab:effectiveness}
% line height
\renewcommand{\arraystretch}{1.2}
\begin{tabular}{clcc|cc|cc|cc|cc|cc}
\toprule
                            &         & \multicolumn{2}{c}{ACC\#1} & \multicolumn{2}{c}{ACC\#2} & \multicolumn{2}{c}{AFC\#1} & \multicolumn{2}{c}{AFC\#2} & \multicolumn{2}{c}{WT\#1} & \multicolumn{2}{c}{WT\#2} \\
                            &         & train        & test        & train        & test        & train        & test        & train        & test        & train       & test        & train       & test        \\\hline 
\multicolumn{2}{c}{Original}            & 96.8\%       & 97.4\%      & 82.0\%       & 84.1\%      & 82.9\%       & 85.4\%      & 86.6\%       & 85.5\%      & 84.0\%      & 84.0\%      & 98.1\%      & 97.7\%      \\
\multirow{3}{*}{\rotatebox{0}{Repaired}
} & \minsat  & 75.6\%       & 76.0\%      & \tbgray 95.4\%       &\tbgray 96.7\%      & 82.3\%       & 83.7\%      & \tbgray 91.5\%       &\tbgray 91.9\%      & 86.1\%      & 85.8\%      & 98.2\%      & 97.8\%      \\
                            & \similarity & \tbgray 97.8\%       & \tbgray 97.9\%      & 81.0\%       & 81.9\%      & 96.0\%       & 96.3\%      & 81.1\%       & 81.5\%      & 86.8\%      & 85.9\%      & 98.2\%      & 97.8\%      \\
                            & \randoms  & 96.3\%       & 97.1\%      & 71.9\%       & 72.3\%      & \tbgray 99.4\%       &\tbgray  99.6\%      & 90.5\%       & 90.3\%      & \tbgray93.1\%      &\tbgray 92.5\%      & \tbgray98.5\%      & \tbgray 98.1\%     \\
                            \bottomrule
\end{tabular}
\end{table*}

\section{Evaluation Results}\label{sec:experimentResults}

In this section, we present our evaluation results used to evaluate our proposed framework \autorepair. 

\medskip
\noindent\textbf{RQ1: Does \autorepair effectively improve the safety of DNN-enabled CPS?}

This RQ aims to evaluate whether \autorepair can truly improve the safety of these subject CPS. We run these subject CPS over the retraining dataset and the test set, with the original controller and with the repaired controller respectively, by feeding the input signals from the executions in the retraining dataset and the test set into the systems. 

The experimental results for RQ1 are shown in Table~\ref{tab:effectiveness}, in which the safety rates of these systems with the original DNN controllers and with the repaired DNN controllers are compared. We notice that, in most of the cases, compared to the relatively low safety rates of the original systems, the systems with the repaired controllers, under different configurations, can achieve significantly higher safety rates. For instance, the safety rates of these systems on the retraining sets increase by at least 0.4$\%$ (for WT\#2) and at most 16.5$\%$ (for AFC\#1); the safety rates of these systems on the test sets increase by at least 0.4$\%$ (for WT\#2) and at most 14.2$\%$ (for AFC\#1). These results evidently exhibit the advantages of \autorepair in terms of safety, in that it indeed repairs the vulnerable characteristics of the original systems successfully. Moreover, the safety improvement on the test set also demonstrates the generalization ability of \autorepair. 

We also notice the cases when the safety rate decreases after repair, e.g., ACC\#1 under \minsat. This issue can be caused by several factors in the approach, including the quality of retraining. In practice, we can mitigate this issue by tuning the hyperparameters in the controller retraining and trying \autorepair with other strategies.

\begin{tcolorbox}[size=title, colback=gray!10, breakable]
{\textbf{Answer to RQ1:}
 \autorepair is able to repair unsafe DNN controllers effectively, and thus increase the safety and reliability of the entire system. }
\end{tcolorbox}

\medskip
\noindent\textbf{RQ2: How does the optimum selection strategy influence the effectiveness of \autorepair?}

In RQ2, we study how different optimum selection strategies of the Pareto front in the signal repair phase (Section~\ref{sec:signalRepair}) influence the effectiveness of \autorepair. In this experiment, we compare the performances of \autorepair under the three optimum selection strategies introduced in Section~\ref{sec:signalRepair}. 

As shown in Table~\ref{tab:effectiveness}, \minsat  outperforms the original controller in 4/6 cases, but is not effective for ACC\#1; \randoms outperforms the original controller in 4/6 cases, and outperforms the other two strategies in 3/6 cases, but is not effective for ACC\#2. These results show that  \minsat and \randoms are effective in general, but they may also be not effective in some cases. As we analyzed in RQ1, in practice, we can improve the effectiveness of \autorepair by trying to improve the quality of retraining and trying different strategies, such that the repaired systems can achieve higher safety rates.

Moreover,
\similarity outperforms the original controller in 4/6 cases, and in the remaining 2 cases, it performs similarly to the original controller; however, only in 1/6 case it outperforms other two strategies. These results show that \similarity is relatively weak in changing the characteristics of DNN controllers, because it always prefers the control signal patchs that have the least deviation from the original control signals.

\begin{tcolorbox}[size=title, colback=gray!10, breakable]
{\textbf{Answer to RQ2:}
Among the three optimum selection strategies, \minsat and \randoms are effective in most cases; \similarity is also effective in many cases, but it is weak in changing the characteristics of DNN controllers.
}
\end{tcolorbox}

\medskip
\noindent\textbf{RQ3: Does \autorepair affect the performance of DNN-enabled CPS in terms of other metrics than safety?}

\begin{figure}[!tb]
\centering  \includegraphics[width=\linewidth]{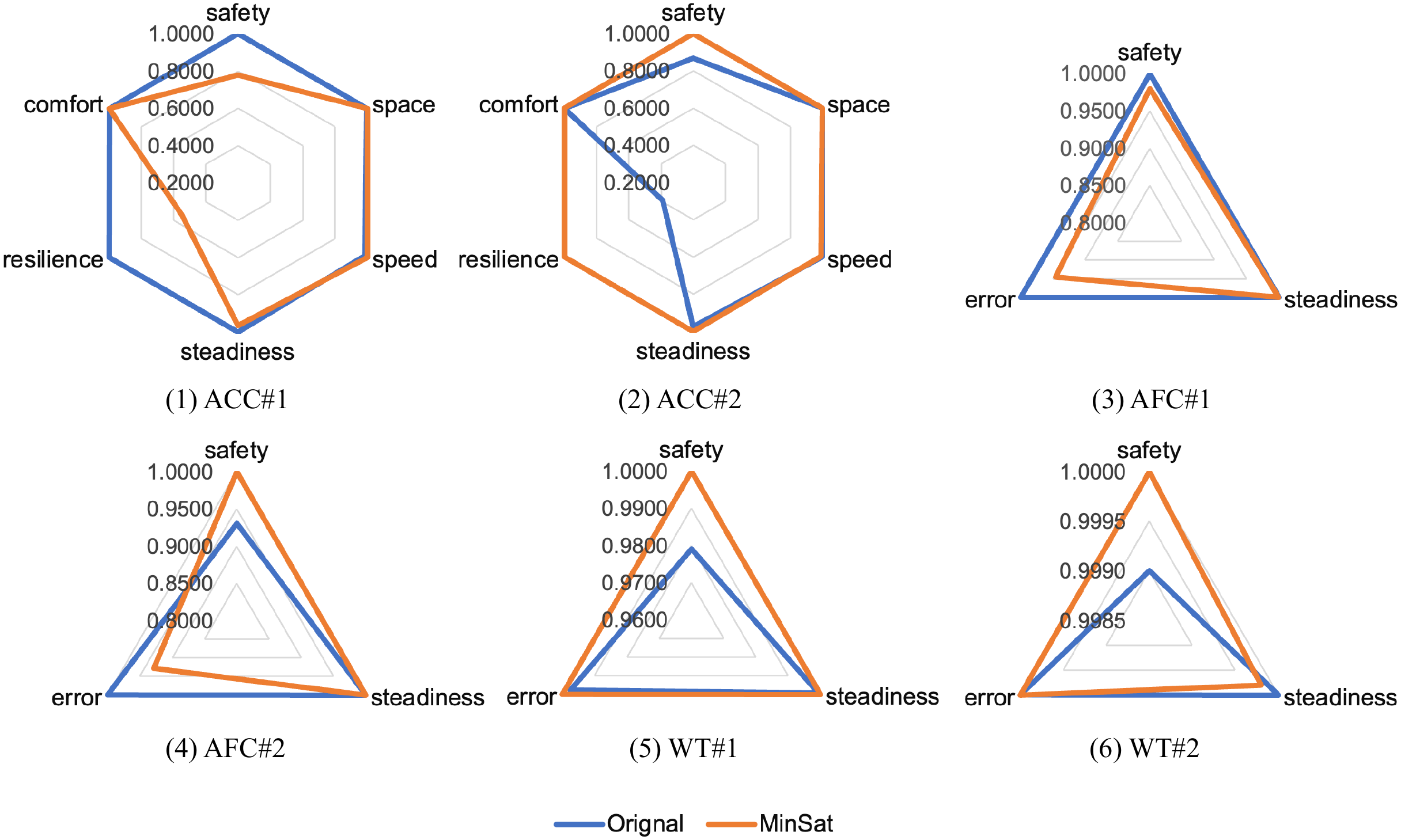}
\caption{Affection of \autorepair to system performances. (RQ3)}
\label{fig:rq3}
\end{figure}

The affection of \autorepair to the performances in different aspects of the systems is shown in Fig.~\ref{fig:rq3}. Specifically, we normalize the performance data in terms of different metrics, and show the performance differences between the original systems and the repaired systems. The blue line and the orange line represent the normalized performances of the original systems and the repaired systems, respectively.

As illustrated in Fig.~\ref{fig:rq3}, compared to the original systems, the repaired systems do not suffer much from performance degradation. For example, in ACC\#2, while \autorepair improves the safety of the original system, it barely harms the space performance (i.e., how far the ego travels) or the speed performance (i.e., the speed should not be too low in a safe status). This is due to that in our repair algorithm, we minimize the deviation of the control signal patch from the original control signal. Of course it also happens that the safety improvement comes with performance degradation in some aspects (e.g., error performance for AFC\#2). The severity of such degradation depends on the importance of these metrics.

Moreover, we observe that \autorepair can also improve the performance in some other aspects, together with safety, e.g., the steadiness for WT\#1. That is because safety and the other performance metrics are consistent in some cases. For instance, the steadiness metric requires the system to stay in a safe status for a certain period, so improving safety can also improve the system performance in steadiness. 

\begin{tcolorbox}[size=title, colback=gray!10, breakable]
{\textbf{Answer to RQ3:} 
\autorepair can improve the safety of the systems with little sacrifice in performance of other aspects. Sometimes, it can even improve the performance in the aspect that is relevant to safety.
}
\end{tcolorbox}

\begin{table}[!tb]
\centering
\caption{The time cost of \autorepair (Time in seconds)}
\label{tab:timecost}
% line height
\renewcommand{\arraystretch}{1.2}
%\resizebox{0.9\linewidth}{!}{
\begin{tabular}{lccrc}
\toprule
       & sim. &      avg. cost    & vio. ratio & total cost \\ \hline
ACC\#1 &    1.42             &          393.81             &   318/10000    &   $1.25 \times 10^5$         \\ 
ACC\#2 &    2.08             &          558.92             &  1800/10000    &   $ 1.01 \times 10^6$         \\  
AFC\#1 &    2.03             &          480.69             &  1713/10000    &   $ 8.23 \times 10^5$         \\  
AFC\#2 &    2.54             &          631.66             &  1339/10000    &   $ 8.46 \times 10^5$         \\  
WT\#1  &    1.07             &          171.77             &  1596/10000    &   $ 2.74 \times 10^5$         \\  
WT\#2  &    1.36             &          262.83             &   192/10000    &   $ 5.05 \times 10^4$         \\ \bottomrule
\end{tabular}
%}
\end{table}

\medskip
\noindent
\textbf{RQ4: How is the efficiency of \autorepair?}

Table~\ref{tab:timecost} lists the  details of the time costs of \autorepair, including the time cost of one system execution, the average cost of repairing one unsafe execution, the ratio of the unsafe executions in the retraining set, and the total time cost for repairing all the unsafe traces in the retraining set.

First, as shown in Table~\ref{tab:timecost}, the cost of repairing a single system execution depends on the time cost of a single system execution, and moreover, it depends on  the configuration of the multi-objective optimization algorithm. The parameter settings of NSGA-II are given in Table~\ref{tab:nsga-ii}; by tuning those parameters, users can balance the trade-off between efficiency and repair effectiveness. 
Second, the total time cost of \autorepair also depends on the ratio of the violating system executions in the collected dataset for retraining. If the system is relatively safe, less repair efforts will be paid. To sum up, as our approach can effectively improve the safety of the DNN-enabled CPS,  paying such repair efforts is worthwhile.

\begin{tcolorbox}[size=title, colback=gray!10, breakable]
{\textbf{Answer to RQ4:} 
The time cost of \autorepair depends on the system execution time, the hyperparameters of optimization algorithms, and the safety status of the systems. For safety-critical systems, it is worth paying such repair efforts to achieve higher safety rates.
}
\end{tcolorbox}

\section{Related Work}

\medskip
\noindent{\bf DNN controllers.} With the rapid development of AI, practitioners consider replacing traditional controllers with DNN controllers~\cite{liu2021recurrent, deshmukh2019learning, yaghoubi2020training, nivison2018development, gilra2018non, ferlez2020two}. A DNN controller can be trained in different ways, including \emph{supervised learning}~\cite{hertneck2018learning} and \emph{deep reinforcement learning (DRL)}~\cite{arulkumaran2017deep}. Supervised learning  requires to collect system executions of an original CPS to train new DNN controllers. For example, Hertneck et al.~\cite{hertneck2018learning} use DNNs to approximate a robust MPC controller. In contrast, DRL follows the \emph{``trial-and-error''} paradigm, learning the best strategy via interactions with environment. A recent work~\cite{Song2021WhenCS}  applies DRL for DNN controller training.

\medskip\noindent{\bf DNN repair.} Different repair mechanisms have been adopted to improve the performance and reliability of DNN models. Existing work on DNN repair can be roughly categorized into two lines---{\em adversarial training} and {\em parameter alteration}. 

{\em Adversarial training} refers to the approaches that first collect adversarial examples and then use these examples with corrected labels to retrain the model. It has been extensively studied in repairing image classifiers, e.g.,~\cite{goodfellow2014explaining, wong2018provable, shafahi2019adversarial, ma2018mode, yu2021deeprepair, gao2020sensei}. While our work could be treated as a specific practice of adversarial training, we differ from existing works in the way we obtain the labels for the adversarial examples, namely, due to the high complexity of DNN-enabled CPS, we search for the corrections for the control signals by optimization, such that safety is achieved at the system level.

{\em Parameter alteration} refers to the approaches that directly modify DNNs' parameters to change the inference of adversarial examples~\cite{sohn2019search, wang2019repairing, zhang2019apricot}. For instances, \cite{sohn2019search, zhang2019apricot} are the early attempts that localize the suspicious DNN parameters and then optimize them for DNN repair. Our work is orthogonal to these works since we do not manipulate DNNs' weights directly.

\medskip
\noindent{\bf Repair of DNN controllers.}
There have been several attempts to repair DNN controllers. Yang et al.~\cite{yang2022neural} propose to first attack a DNN controller and then retrain with the identified adversarial examples. However, compared to~\cite{yang2022neural} in which an isolated controller is considered, we target at a physical system controlled by DNN controller; also, our safety requirement is defined on the system level. Another related work is~\cite{zhou2020runtime}, in which Zhou et al. repair a learning-based controller using an assumptive \emph{safe} oracle. Compared to~\cite{zhou2020runtime}, our approach automates the controller repair process, by searching in the control decision space for corrections.

\medskip
\noindent{\bf Safe reinforcement learning.}
\emph{Reinforcement learning (RL)} is a typical approach for training controllers. 
In RL, there is a line of work~\cite{alshiekh2018safe, anderson2020neurosymbolic, zhu2019inductive, berkenkamp2017safe} that aims to ensure the safety of the trained controllers; in particular, \emph{shielding}~\cite{alshiekh2018safe} provides a similar mechanism to our framework, which repairs the actions that trigger unsafe system behavior. However, there are several major differences between safe RL and our framework. First, many of the problem settings in RL are bounded to \emph{Markov decision processes (MDP)}, which involve finitely many states and transitions; in contrast, we consider the control of non-linear continuous plants with dense-time real-valued dynamics, which are intrinsically hard to verify. Moreover, the safety mechanisms like shielding can be done at runtime---when an unsafe action is detected, a shield can be activated to stop it; in our case, banning a single action at runtime does not prevent unsafety from happening, due to the systems' \emph{dense-time} nature, and this can also explain why we analyze and repair complete signals in an offline manner.

\section{Conclusion and Future Work}
While DNN-enabled CPS have received increasing attention, how to enhance DNN controllers remains an unaddressed challenge. In this work, we propose \autorepair, a general automated repair framework that accounts for system-level safety requirements during repairing. \autorepair adopts a novel signal diagnosis and repair method that, given a system execution trace, identifies the minimal changes to the initial sequence of control signals to avoid unsafe system outputs. By applying the optimal changes to the control sequence, \autorepair obtains a new safety-assured execution trace, which can be used to repair the initial DNN controller. Our experiments show that \autorepair can effectively repair unsafe DNN controllers and increase the safety of the entire system. In future, we plan to extend our framework to support the safety requirements specified by more diverse advanced temporal logic formalism towards providing better safety assurance and enhancement to accelerate the DNN adoption in diverse industrial CPS across domains.

\section*{Acknowledgments}
This work was supported in part by funding from the Canada First Research Excellence Fund as part of the University of Alberta’s Future Energy Systems research initiative, Canada CIFAR AI Chairs Program, Amii RAP program, the Natural Sciences and Engineering Research Council of Canada (NSERC No.RGPIN-2021-02549, No.RGPAS-2021-00034, No.DGECR-2021-00019), as well as JSPS KAKENHI Grant No.JP19H04086, No.JP20H04168, No.JP21H04877, JST-Mirai Program Grant No.JPMJMI20B8. D. Lyu is also supported by JST SPRING Grant No. JPMJSP2136. 

\bibliographystyle{IEEEtran}
\bibliography{reference}

\begin{comment}

\end{comment}

\end{document}